# Mode Coupling in a Tapered Slow Light Waveguide

S. He, Y. He, Y. Jin, and J. He
COER, Zhejiang University, China

**Metamaterials [1-4] with simultaneous negative permittivity and negative permeability (also called left-handed materials) open new avenues to achieving unprecedented physical properties and functionality unattainable with naturally occurring materials. It has been predicted that a metamaterial slab waveguide can slow down light significantly (even to zero velocity) as the thickness of the core layer approaches a critical thickness [5, 6]. Here we show that coupling between the forward and backward modes becomes significant near the critical thickness due to the good matching of the wavevectors and modal profiles of the two modes. Physical explanation and impact of this coupling are given.**

In the metamaterial slab waveguide consisting of a left-handed material (LHM, in which the wave vector and energy flux are in opposite directions) core and right-handed material (RHM, i.e., conventional dielectric) cladding, the forward and backward (meaning the energy flux velocity $v_E$ and wave vector are in opposite directions) modes coexist (see the upper and lower branches of the fundamental oscillating $TM_2$ mode in Fig. 1a below), and the propagation constants of a right-going (or left-going) forward mode and a left-going (or right-going) backward mode degenerate (i.e., $\beta_f = \beta_b$) at the critical core thickness (denoted by $d_c$; see Fig. 1a below). In this work we use mode matching technique [7] and the finite element method (FEM) to investigate the performance of slow light waveguide with an adiabatically varying core layer thickness.

In all our numerical studies, the tapered waveguide has the following parameters: $\varepsilon_1 = \mu_1 = 1 - \omega_p^2/(\omega_p^2 + i\omega\Gamma)$ (for the LHM core) with $\omega_p = \sqrt{6}\omega_c$, $\omega_c/2\pi = 1$ $THz$ (corresponding to $\lambda_c = 300$ $\mu m$), $\varepsilon_2 = 2.56$ (relative permittivity for lower cladding), $\mu_2 = 1$, $\varepsilon_3 = 2.25$ (for upper cladding), $\mu_3 = 1$. All the following calculations are for $TM$ polarized monochromatic continuous wave at the frequency of 1 THz except as otherwise indicated. Here we assume the metamaterial is lossless, i.e., $\Gamma = 0$.

Fig. 1a and b show how the propagation constant $\beta$ and energy flux (group) velocity $v_E$ vary with core-thickness $d$ for an LHM slab waveguide (see the inset in Fig. 1a). Here one sees clearly a critical core thickness $d_c$, which separates the propagation region ($d > d_c$) and the attenuation region ($d < d_c$). The critical thickness $d_c$ is calculated to be *52.451813091371 μm* for fundamental oscillating $TM_2$ mode. At the critical thickness, the counter-directional energy fluxes balance each other and result in a slow waveguide with zero group velocity. A forward guided mode has more energy confined in the RHM cladding, while a backward guided mode has more energy confined in the LHM core layer (see the inset for point $O_1$ in Fig. 1c).

Consider a waveguide with a continuously tapered thickness for the LHM core and assume an incident forward mode is excited from the left. Here the left-going backward guided mode with negative group velocity has positive wave vector, which is in the same direction as the wave vector of the incident forward guided mode. As $d$ approaches $d_c$, the incident right-going forward mode (see point $M_f$ on the lower branch in Fig. 1a) and the left-going backward mode (see point $M_b$ on the upper branch of the same order in Fig. 1a) get similar wave vectors (similar magnitude and same sign). It therefore fulfills the requirement of wavevector (momentum) matching for coupling between the forward and backward guided modes. It is well-known that wavevector (momentum) matching will give a large probability for e.g. three-wave mixing process (the so-called 'condition of perfect phase matching') [8] and radiative recombination in direct band gap semiconductor (the so-called 'conservation of momentum or total k-vector') [9], and small detuning from the matching condition will reduce the occurrence probability of the event/process. Similarly, coupling between forward and backward modes is more likely to occur when the wavevector (momentum) mismatch between the two modes decreases (i.e., when $d$ approaches $d_c$) in a continuously-tapered waveguide (without any sudden material change along the propagation direction; these wavevectors are tangential to the two LHM/RHM boundaries of the waveguide). Efficient coupling between two modes also requires a good matching of the two modal profiles. As $d$ approaches $d_c$, the modal profiles of the backward and forward modes become very similar (see the inset for point $O_2$ in Fig. 1c), and the value of mode overlap between the two modal profiles gets closer to unity (see the overlap curve in Fig. 1c).

In our study, we consider the case when the core-thickness $d$ is not large so that only one oscillating mode, namely, $TM_2$ mode, can be supported in the metamaterial waveguide. Here we calculate the reflection of optical power as the LHM core thickness of the output waveguide decreases gradually to critical thickness $d_c$. We use a reliable mode analysis and matching technique (see e.g. [7]) to calculate the reflection due to the coupling between the forward and backward modes. It is difficult to obtain an analytical field expression for a waveguide of continuously varying shape. Instead, we use a series of steps to approximate the tapered section so that the mode matching technique can be employed to obtain the reflection and transmission of optical energy in a semi-analytic way. We analyze separately for the cases of $d_{out} < d_c$ and $d_{out} > d_c$.

For the case of $d_{out} < d_c$, the output waveguide cannot support any guided mode, and consequently no transmission is expected. We approximate the tapered structure with many steps and increase the total number of steps $N$ to approach a smoothly tapered waveguide (see Fig. 2a). The incident optical power carried by a right-going forward mode in the input waveguide will be channeled to different modes after it enters the tapered waveguide. The exact ratio of the optical power in each mode is calculated with the mode matching technique and shown in Fig. 2b. In particular, we are interested in the following power ratios (see the colored arrows in Fig. 2a): $R_{11}$ (ordinary reflection from the incident right-going forward mode to the left-going forward mode), $R_{12}$ (reflection due to the coupling from the incident right-going forward mode on the lower branch to the left-going backward mode on the upper branch, see the red open arrow in the inset of Fig. 2a) and the power scattered to leaky modes. Our calculation results of Fig. 2b show that, (1) when the number of steps $N$ is small (~10), the optical power scattered into leaky modes is observable. However, it deceases quickly as the tapered waveguide becomes smoother and smoother ($N>20$); (2) when $N$ is small, $R_{11}$ is comparable with $R_{12}$. However, $R_{11}$ also deceases to nearly zero as the tapered waveguide becomes smoother and smoother ($N>60$); (3) for a smooth enough waveguide, $R_{12}$ is precisely unity, which implies that all the energy of the incident right-going forward mode will be eventually coupled to the left-going backward mode and flow out from the input port.

Then we study the case of $d_{out} > d_c$. In this case (see Fig. 2c), the output waveguide still supports some guided modes. Fig. 2d shows the calculated $R_{12}(0, z)$ (accumulated contribution to reflection $R_{12}$ from the coupling in partial section $[0, z]$ of the whole tapered waveguide) and $T_{11}(0, z)$ as $z$ increases. $R_{11}$ is zero because the tapered waveguide is smooth enough. From Fig. 2d one sees that more and more incident optical power is reflected as $z$ increases. The starting section [where $d_{gap}$ ($\equiv d-d_c$) is large (*1* or *2* $\mu m$)] of the tapered waveguide contributes little to the reflection. However, the reflection increases significantly as $d_{gap}$ approaches 0 (i.e., when $z$ is close to the critical thickness). In the end, the reflection gets saturated to unity. All the incoming energy is coupled to the backward mode before the group velocity decreases to exactly zero. Clearly, the incident optical power suffers strong reflection (due to the coupling) when $d_{gap}$ is small *($d_{gap}<10^{-3}$ $\mu m$)*. Fig. 2e shows the case of a 10-time longer tapered section. The solid curves represent the optical power ratios when the incident power is carried by a right-going forward mode, and the dashed curves with stars represent the optical power ratios when the incident power is carried by a right-going backward mode. As expected, they give exactly the same curves (due to the well-known reciprocity theory). Compared with Fig. 2d, one sees that a 10-time longer tapered waveguide gives the same trend, and cannot help in reducing reflection $R_{12}$.

Fig. 2f shows the distribution of the magnetic field component $|H_y|$ calculated with the mode matching technique in a tapered metamaterial waveguide. To validate the mode matching technique, we also calculate the field distribution with FEM (finite element method) of commercial software COMSOL. The FEM simulation gives exactly the same field distribution pattern as the calculation of the mode matching technique. Fig. 2g gives a comparison of the field distribution along line *x=0*. The two lines fit each other very well, which shows the reliability of our results calculated with the mode matching technique. From this figure we see clearly some standing-wave characteristics in field magnitude $|H_y|$ (normalized with the incident field magnitude), with alternative peaks and nodes along the *z* direction. This is also a clear evidence for the excitation of left-going backward mode: the backward mode and forward mode interfere with each other, forming alternative peaks and nodes of the field magnitude. From Fig. 2g one also sees that the peak value grows gradually as they get closer to the critical thickness, reaching a maximum of over *3.75*. This local field enhancement around the critical thickness results from the slow light effect.

In summary, we have found the coupling between the forward and backward modes in a tapered slow light waveguide and given a clear physical explanation. Its impact to the reflection has also been studied.

**References**

[1] J. B. Pendry, "Negative Refraction Makes a Perfect Lens." Phys. Rev. Lett. **85**, 3966 (2000).
[2] Shelby, R. A., Smith D. R., et al. "Experimental Verification of a Negative Index of Refraction." Science **292**, 77-79 (2001).
[3] Pendry J. B., Schurig D., & Smith D. R., "Controlling Electromagnetic Fields." Science **312**, 1780 – 1782 (2006).
[4] Valentine, J., Zhang, S., et al., "Three-Dimensional Optical Metamaterial Exhibiting Negative Refractive Index", Nature **455**, 376-340 (2008).
[5] He, J. & He, S. "Slow propagation of electromagnetic waves in a dielectric slab waveguide with a left-handed material substrate." IEEE Microwave and Wireless Components Letters, **16**, 96-98 (2005).
[6] Tsakmakidis, K. L., Boardman, A. D. & Hess, O. "`Trapped rainbow' storage of light in metamaterials." Nature **450**, 397-401 (2007).
[7] Reiter, J. M. & Arndt, F. "Rigorous Analysis of Arbitrarily Shaped H- and E-Plane Discontinuities in Rectangular Waveguides by a Full-Wave Boundary Contour Mode-Matching Method." IEEE Trans. Microwave Theory Techn. **43**, 796-801 (1995).
[8] Robert W. Boyd, "Nonlinear Optics" Section 2.2 (Academic, San Diego, 2003, 2nd Ed.).
[9] see e.g. Wikipedia: http://en.wikipedia.org/wiki/Direct_and_indirect_band_gaps
[10] He, S. & Strom, S. "The electromagnetic scattering problem in the time domain for a dissipative slab and a point source using invariant imbedding." J. Math. Phys. **32**, 3529-3539 (1991).
[11] Loh, P. R., Oskooi, A. F., et al. "Fundamental relation between phase and group velocity, and application to the failure of perfectly matched layers in backward-wave structures." Phys. Rev. E **79**, 065601 (2009).

**FIGURE CAPTIONS**

**Figure 1. Analysis of forward and backward *TM$_2$* modes as the core thickness (*d*) of a metamaterial slab waveguide varies.** (**a**) The propagation constants of the forward and backward *TM$_2$* modes (at the frequency of 1THz) as *d* varies. (**b**) Energy flux velocity *v$_E$* as core thickness *d* varies. Energy flux velocity is defined by *v$_E$/c$\equiv$ (P$_1$+P$_2$+P$_3$)/(|P$_1$| + |P$_2$| + |P$_3$|)*, where *P$_i$ (i=1, 2, 3)* denotes the optical

power flow in the *i-th* layer. (**c**) Overlap of the forward and backward guided modes as LHM core thickness *d* varies. The mode overlap is defined as

$$Overlap \equiv \frac{|\int H_f H_b dx|^2}{\int |H_f|^2 dx \int |H_b|^2 dx},$$ where $H_f$ (in pink color) and $H_b$ (in aquamarine blue) denote $H_y$ components of the forward and backward modes, respectively. Inset: mode profiles of $H_f$ and $H_b$ for two different core thicknesses, namely, $d = 54.7 \mu m$ at point $O_1$, and $d = 52.7 \mu m$ at point $O_2$ (very close to critical thickness $d_c \approx 52.45 \mu m$).

**Figure 2. Reflection from a tapered metamaterial waveguide which can support only $TM_2$ mode.** (**a**) Reflection and transmission of different types for the case of $d_{out} < d_c$. The thick open arrows show the direction of the energy flux, while the thin arrows (inside the thick open arrows) indicate the direction of wave vector. (**b**) Optical power ratio for each type of reflection or leakage, as the total number of steps increases. Here we choose $d_{in} = 55 \mu m$, $d_{out} = 52.4 \mu m$ ($<d_c$), and the length of the tapered section is fixed to $L_{taper} = 20 \lambda_c$. (**c**) A case with $d_{out} > d_c$. In order to give a deep physical insight to the reflection, we use the concept of invariant imbedding (see e.g. [10] and relevant references therein) to study the contribution of each slice of the tapered waveguide to the reflection. Specifically, we choose two planes perpendicular to the *z* axis in the tapered waveguide, denoted by Plane A ($z=z_1$) and Plane B ($z=z_2$). Then we can calculate $r_{12}(z_1, z_2)$, $r_{11}(z_1, z_2)$, and $t_{12}(z_1, z_2)$ as $z_1$ or $z_2$ varies. For example, $r_{12}(0, 20\lambda_c)$ is the physical reflection for the whole tapered waveguide while $r_{12}(0, z)$ tells the accumulated contribution from the coupling in partial section $[0, z]$ of the whole tapered waveguide [one may derive an imbedding differential equation for $r_{12}(0, z)$ with boundary condition of $r_{12}(0, 0) = 0$]. The reflection (transmission) of optical power can be obtained simply by $R_{ij} = |r_{ij}|^2$, $T_{ij} = |t_{ij}|^2$, ... $i,j=1,2$. (**d**) and (**e**) Accumulated contribution to the reflection and transmission from segment $[0, x]$ of the whole tapered waveguide with $d_{out} > d_c$. The length of the tapered waveguide is $50\lambda_c$ and $500\lambda_c$ for **d** and **e**, respectively. Here we choose $d_{in} = 55 \mu m$, and $d_{gap}(z)=(d(z)-d_c)*10^{-kz}$, where tapering coefficient $k=7.3*10^{-4}/\mu m$ for **d** (in order to obtain a slowly-varying waveguide), and $k = 7.3*10^{-5}/\mu m$ for **e** (to ensure the same core thickness at the output waveguide). (**f**) Distribution of field magnitude $|H_y|$ calculated with the mode matching technique. The tapered section is located between $z=0$ and $z=6000 \mu m$. The LHM core thicknesses of the input and output waveguides are $55 \mu m$ and $52.4 \mu m$, respectively. (**g**) show the plots of $|H_y|$ calculated with two different methods along line $x=0$. In the FEM simulation, an adiabatic scalar absorbing layer [11] is used to eliminate parasitic reflection, and the forward mode is excited with a profile-specified source.

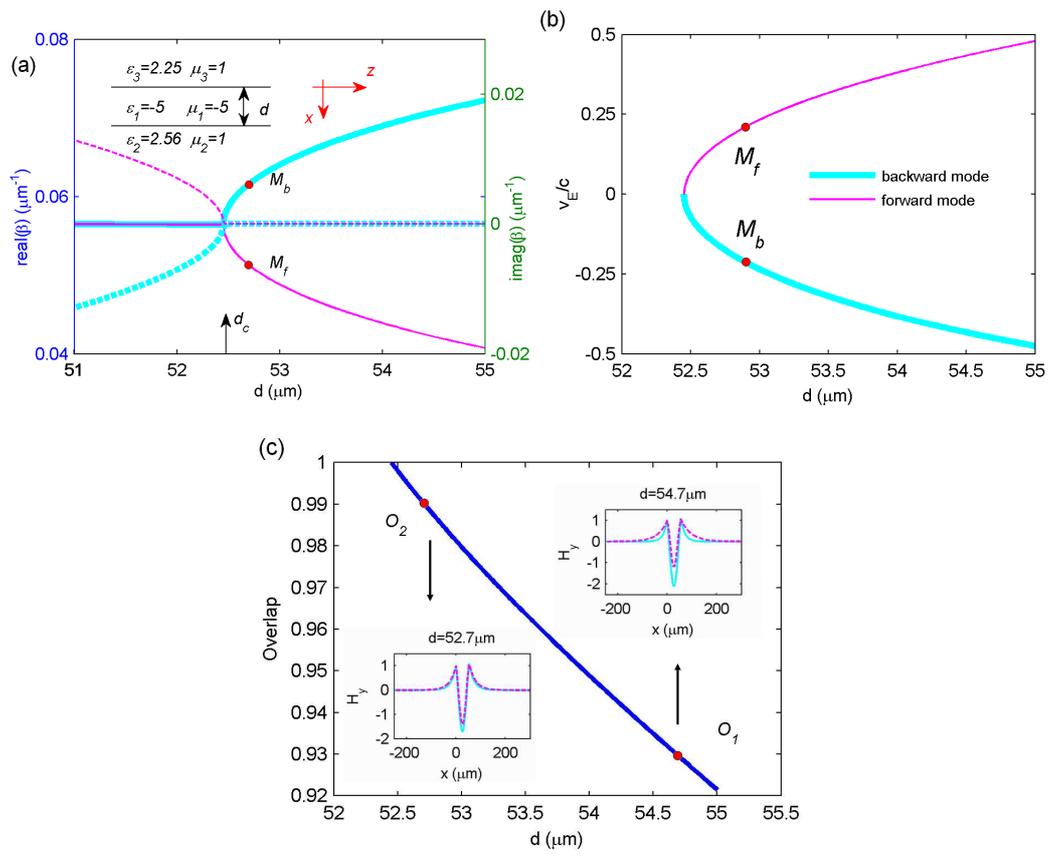

Figure 1.

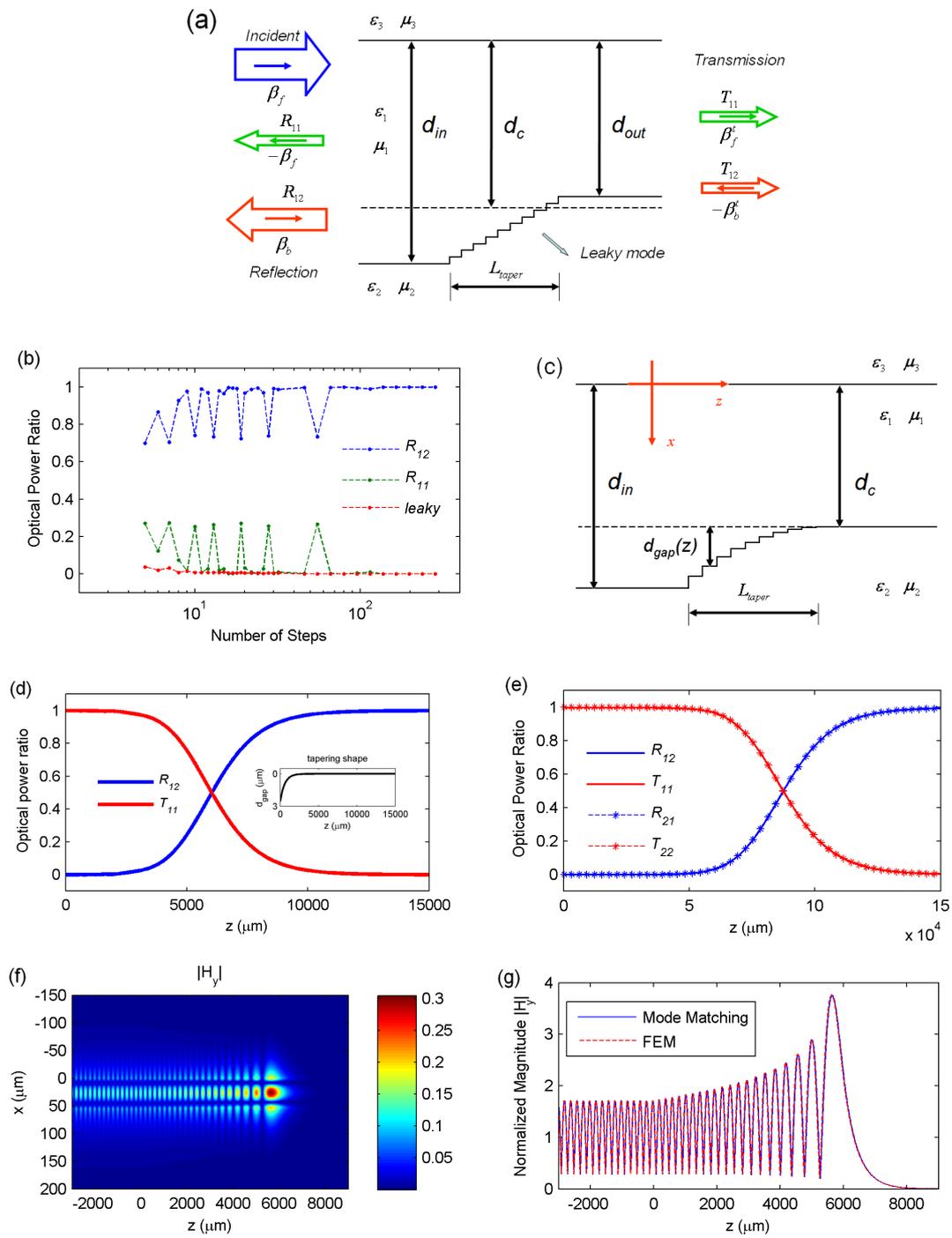

Figure 2.